\documentclass[twocolumn,aps]{revtex4}
\usepackage{graphicx,amssymb}

\begin{document}

\title{Coherent acoustic vibration of metal nanoshells}

\author{C. Guillon$^a$, P. Langot$^a$, N. Del Fatti$^{a,b}$, and F. Vall\'ee$^{a,b}$}
\affiliation{$^a$ Centre de Physique Mol\'eculaire Optique et Hertzienne\\
CNRS and Universit\'e Bordeaux I, 351 cours de la Lib\'eration, 33405 Talence, France}

\affiliation{$^b$ Laboratoire de Spectrom\'etrie Ionique et Mol\'eculaire\\
CNRS and Universit\'e Lyon I, 43 Bd. du 11 Novembre 1918, 69622 Villeurbanne, France}

\author{A. S. Kirakosyan and T. V. Shahbazyan}
\affiliation{Department of Physics and Computational Center for Molecular Structure and Interactions, Jackson State University,
P.O. Box 17660, Jackson, Mississippi 39217, USA}

\author{T. Cardinal and M. Treguer}
\affiliation{Institut de Chimie de la Mati\`ere Condens\'ee de Bordeaux\\
CNRS and Universit\'e Bordeaux I, 87 Av. du Dr. Albert Schweitzer, 33608 Pessac, France}

\date{\today}

\begin{abstract}
Using time-resolved pump-probe spectroscopy we have performed the
first investigation of the vibrational modes of gold nanoshells. The
fundamental isotropic mode launched by a femtosecond pump pulse
manifests itself in a pronounced time-domain modulation of the
differential transmission probed at the frequency of nanoshell surface 
plasmon resonance. The modulation amplitude is significantly stronger
and the period is longer than in a gold nanoparticle of the same
overall size, in agreement with theoretical calculations. This
distinct acoustical signature of nanoshells provides a new and 
efficient method for identifying these versatile nanostructures and
for studying their mechanical and structural properties.
\end{abstract}

\maketitle

Metal nanoshells -- metal shells grown on dielectric spheres -- are
among the highlights of nanostructures with versatile optical and 
mechanical properties \cite{halas-prl97}. As for fully metallic
nanoparticles, absorption and scattering of light by nanoshells are
dominated by the surface plasmon resonance (SPR)
\cite{kreibig}. However, they offer wide new possibilities of 
controlling the SPR characteristics, such as its frequency position,
by varying, for instance, the shell thickness vs. overall size, or the
constituent materials \cite{halas-prb02,prodan-nl03}. Furthermore,
recent measurements of scattering spectra  of single nanoshells
\cite{klar-nl04,halas-nl04} indicated enhanced sensitivity to their
environment and narrowing of their resonance lineshape as compared to 
solid gold particles \cite{klar-prl98}. These unique tunability and
characteristics of nanoshells optical properties spurred a number of
proposals for their applications including in optomechanics
\cite{halas-ap01}, sensing \cite{sun-ac02,halas-apl03}, and drug
delivery \cite{halas-ac03,halas-nl05}.

While the optical response of nanoshells has been extensively studied,
much less is known about their acoustical properties. In fact, the 
low frequency vibrational modes of nanostructures bear a unique
signature of their structural and mechanical properties directly
reflecting the impact of confinement on the ionic movement. This is in
contrast to the optical frequency domain whose features are determined
by the electronic response. The vibratinal modes  thus constitute
additional source of information that could be particularly important
in the case of complex systems. This is especially the case for hybrid
or layered systems with structurally distinct constituents, such as
nanoshells \cite{zhang-nl05,klar-nl05}. Due to their structural
sensitivity, vibrational modes are also expected to constitute
acoustic signatures permitting further nanoshell identification,
complementary to the optical one. 
 
Time-resolved optical techniques are powerful tools for investigating
the low-frequency vibrational modes of nanostructures. They have been
applied to the study of the acoustic properties of semiconductor and
metal nanoparticles
\cite{delfatti-jcp99,hartland-jcp99,elsayed-nl04,chergui-nl06} and
have recently been extended to bimetallic particles
\cite{hartland-bimetal1,hartland-bimetal2}. In these experiments,
vibrational modes are impulsively excited as a result of rapid
expansion of metal particle induced by absorption of a femtosecond
pump pulse \cite{delfatti-jcp99,voisin-ss00}. In this process, the
energy initially injected into the electron gas is quickly damped to
the lattice on the time scale of the electron-phonon energy transfer,
about 1-2 picosecond in noble metals. Due to lattice anharmonicity,
this lattice heating leads to an isotropic force on ions triggering
in-phase dilation of each particle that subsequently undergoes radial
contractions and expansions around its new equilibrium size. The
periodic change in nanoparticle volume translates into a modulation in
time of the electronic properties. This can be detected by a
time-delayed probe pulse monitoring the concomitant modulation of the
wavelength of the SPR \cite{delfatti-jcp99}. The triggered initial
expansion being homogeneous, the modulation is dominated by the
fundamental breathing mode that closely corresponds to particle
expansion as a whole. 

Using time-domain spectroscopy, we have performed the first
investigation of acoustic vibrational modes in metal nanoshells. As in
fully metallic particles, we observed in nanoshells a pronounced
time-modulation of the measured probe differential transmission
indicating coherent excitation of the fundamental breathing mode. The
modulation amplitude is significantly stronger and its period is
considerably longer than those in pure gold nanospheres of the same
size. At the same time, the damping time of the oscillations is much
shorter than for gold particles suggesting a faster energy transfer
from nanoshells to the surrounding medium. Such distinct signatures
allow unambiguous identification of nanoshell acoustic vibration and
separation of their contribution from that of possibly present other
entities in a colloidal solution. These results are consistent with
theoretical analysis of vibrational modes in nanoshells.

\begin{figure}[h]
\centering
\includegraphics[width=3.0in]{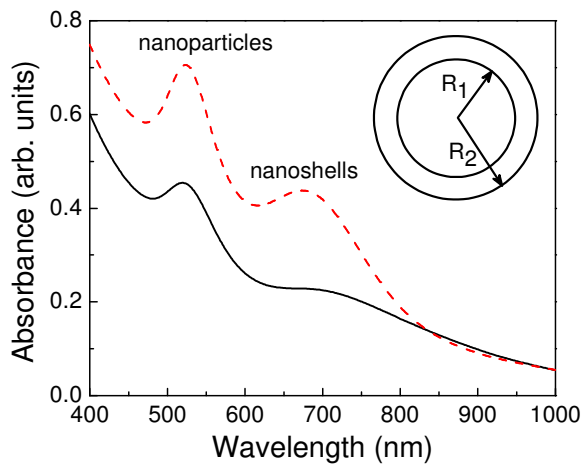}
\caption{\label{fig:absorp}Measured linear absorption spectrum of
  Au$_2$S/Au core/shell particles in water for two different sizes,
  with $R_1=9.0$ nm inner radius and $R_2=11.5$ nm outer radius (solid
  line), and for $R_1=11.5$ nm and $R_2=13.5$ nm (dashed line). Inset:
  schematic geometry of a nanoshell.} 
\end{figure}

Experiments were performed in colloidal solution of
Au$_2$S-core/Au-shell nanoshells prepared using the method described
in Ref.~\cite{halas-prl97}. Both nanoshells and gold nanoparticles are
simultaneously synthesized, as shown by the linear absorption spectra 
of the colloidal solution (Fig.~\ref{fig:absorp}). 
It exhibits two characteristic bands centered around 700 and 530~nm that have been associated to
SPR in core-shell nanoparticles and gold nanospheres, respectively \cite{halas-prl97, aden-jap51}.
This assignement is confirmed by the spectral displacement of the former band during nanoshell growth, while, in contrast, the spectral position of the latter remains almost unchanged as expected for small nanospheres
\cite{halas-prl97}. This is further corroborated by TEM measurements
showing the presence of large nanoparticles (mean radius $R_2$ in the
14 nm range) that have been identified as nanoshells, and of smaller
ones (mean size of about 4 nm) identified as pure gold
\cite{halas-prl97,klar-nl05}. The nanoshell SPR energy is determined
by the ratio of the inner to outer radius $R_1/R_2$; the latter was
thus estimated by fitting the experimental spectrum with 
$A(\omega) =P_p A_{p}(\omega)+ P_s A_{s}(\omega)$, 
where $P_p$ and $P_s$ are the volume fractions of
nanoparticles and nanoshells, respectively, and $A_{p,s}(\omega)$ are
the corresponding absorbances \cite{halas-prl97}. For the investigated
nanoshells, $R_1$ ranges from 9 to 10.3~nm with a shell thickness
$d=R_2-R_1$ of 2.5 to 3.7~nm, i.e., $R_1/R_2$ ranges from 0.78 to 0.73 nm. 

Time resolved measurements were performed using a femtosecond
Ti:sapphire oscillator delivering 20 fs pulses at 860~nm with a 80~MHz 
repetition rate. Part of the pulse train is selected to create the
pump pulses. Absorption in this spectral range being dominated by the 
broad SPR of the nanoshells, they are predominantly excited. The pump
induced transient change of the sample transmission $\Delta T$ is 
detected at the same wavelength around the nanoshell SPR using the
remaining part of the beam. This probe wavelength permits further
selection of the nanoshells that thus dominate the detected $\Delta
T/T$ signal. This has been confirmed by the fact that pure gold 
nanoparticle colloidal solutions yield an undetectable signal in this
pump-probe configuration. The two beams were focussed on a 30~$\mu$m 
diameter focal spot and the pump beam average power was about
100~mW. Measurements were performed with a standard pump-probe setup
with mechanical chopping of the pump-beam and locking detection of $\Delta T$.

\begin{figure}[t]
\centering
\includegraphics[width=3.0in]{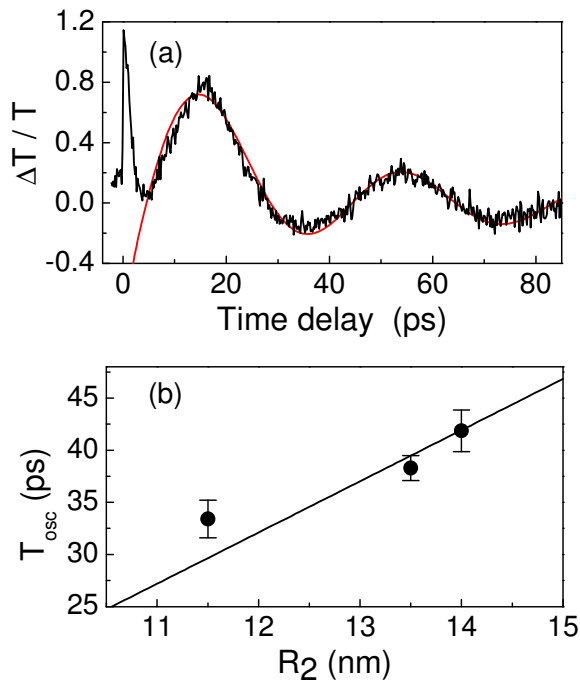}
\caption{\label{fig:oscill} (a). Time-dependent differential
 transmission $\Delta T/T$ measured in  Au$_2$S/Au nanoshells in
 water with near infrared pump and probe pulses is shown together
 with a fit using Eq.~(\ref{fit}).
The inner and outer radii are $R_1=10.3$ and $R_2=13.5$~nm,
 respectively.  (b) Oscillation period $T_{osc}$ measured in different
 Au$_2$S/Au nanoshells as a function of their outer radius $R_2$. The
aspect ratio $R_1/R_2$ is 0.78, 0.76 and 0.74, in increasing $R_2$
order. The line is a fit assuming that $T_{osc}$ is proportional to $R_2$.}
\end{figure}

The measured time-dependent transmission change shows a fast
transient, ascribed to photoexcitation of nonequilibrium electrons and
their cooling via electron-lattice energy transfer
(Fig.~\ref{fig:oscill}). The observed kinetics is consistent with that
previously reported in gold nanoparticles and films for similar
excitation conditions \cite{vallee-jpcb01}. This signal is followed by
pronounced oscillations that can be reproduced by a phenomenological
response function: 
\begin{equation}
\label{fit} R(t) = A \exp (- t/\tau) \cos [2\pi t/T_{osc} - \varphi]\ .
\end{equation}
using a period $T_{osc} \approx 38$~ps and a decay time $\tau \approx
60$~ps for the nanoshells of Fig.~\ref{fig:oscill}(a). Such long
probe-delay response is similar to that reported in fully metallic
nanoparticles
\cite{delfatti-jcp99,hartland-jcp99,elsayed-nl04,chergui-nl06}, but
the measured period of the oscillations is by far too long to ascribe 
them to the residual fully metallic small nanoparticles. We thus
ascribed them to the acoustic vibration of the
nanoshells. Furthermore, their amplitude relative to that of the short
time delay electronic signal, is much larger than in metal nanospheres
(about 75\% as compared to 10\% \cite{delfatti-jcp99}). The measured
period is also much longer (about 4 times) than that predicted for
solid Au nanospheres of the same overall size, about 9~ps for $R_2
=13.5$~nm. Furthermore, the phase of the oscillation 
$\varphi \approx 1.1$ is significantly larger than predicted for a
purely displacive type of excitation in a harmonic oscillator model
(about 0.2). As this phase is a signature of the excitation mechanism,
this suggest a modified launching process as compared to the breathing
mode of nanospheres \cite{voisin-ss00}.

Measurements performed in different nanoshells  show similar behaviors
with an almost linear increase of the oscillation period with the
outer nanoshell size $R_2$ [Fig.~\ref{fig:oscill}(b)]. The effective
decay time $\tau$ of the oscillations can also be extracted from  the
time-domain data. It varies from 30 to 60~ps for the three
investigated samples (30, 60 and 42~ps in increasing order of $R_2$),
but conversely to the period, no systematic variation with the
nanoshell size is experimentally found. This suggests that, as for
nanosphere colloidal solutions, inhomogeneous relaxation  due to the
particle size and structure fluctuations dominates over the
homogeneous one due to matrix-water coupling \cite{hartland-jcp99,voisin-physica02}.

To further correlate the observed oscillation mode with the
nanoparticle structure, we have theoretically analyzed the radial
vibrational modes of a spherical nanoshell in a dielectric medium. The
motion of nanoshell boundaries is determined by the radial
displacement $u(r)$ that satisfies the Helmholtz equation (at zero
angular momentum)
\begin{equation}
u'' + \frac{2u'}{r} + k^2u = 0,
\label{Helm}
\end{equation}
where $k = \omega/c_L$ is the wave-vector. In the presence of core and
outer dielectric medium, the boundary conditions impose that both
the displacement $u(r)$ and the radial diagonal component of the stress tensor,
\begin{equation}
\sigma_{rr} = \rho \biggl[c_L^2 u' + (c_L^2 - 2 c_T^2)\, \frac{2u}{r}
\biggr],
\label{stress}
\end{equation}
are continuous at the core/shell and shell/medium interfaces ($\rho$
and $c_{L,T}$ are, respectively, density and longitudinal/tranverse
sound velocities). In the core, shell, and medium regions, solutions
are, respectively, of the form 
$u \sim \bigl[ \sin(kr)/r\bigr]^{\prime}$,
$u\sim\bigl[\sin(kr+\phi)/r\bigr]^{\prime}$, and   
$u \sim \bigl[e^{ikr}/r\bigr]^{\prime}$, where $\phi$ is the phase
mismatch and prime stands for derivative over $r$.  Matching $u$ and 
$\sigma_{rr}$ at $r=R_1, R_2$ yields the equations for the eigenvalues 
\begin{eqnarray}
\hspace{-2mm}
\frac{\xi^2 \kappa^2}{\xi \kappa \cot(\xi \kappa + \phi) - 1}
-
\frac{\eta_{c}\xi^2 \kappa^2}{(\xi \kappa/\alpha_{c})
\cot(\xi  \kappa/\alpha_c) - 1} +\chi_c =0,
\nonumber \\
\frac{\xi^2}{\xi \cot(\xi + \phi) - 1}
+
\frac{\eta_{m}\xi^2}{1 + i \xi/\alpha_m} + \chi_m = 0,
\qquad \qquad
\label{breathmodes}
\end{eqnarray}
where $\xi=kR_2=\omega R_2/c_L$ and $\kappa  = R_1/R_2$ are shorthand
notations for the normalized eigenenergies and aspect ratio, respectively. The
parameters
\begin{eqnarray}
&&
\alpha_{i} = c_{L}^{(i)}/c_{L}^{(s)},
~~
\eta_{i} = \rho^{(i)}/\rho^{(s)},
~~
\chi_{i} = 4 (\beta_s^2- \eta_i \delta_i^2),
\nonumber \\
&&
\beta_{i}=c_{T}^{(i)}/c_{L}^{(i)},
~~
\delta_{i}=c_{T}^{(i)}/c_{L}^{(s)},
\label{parameters}
\end{eqnarray}
characterize the metal/dielectric interfaces ($i=c,s,m$ stand for
core, shell, and outer medium). From Eq.\  (\ref{breathmodes}), the
ideal case of a nanoshell in vacuum is obtained by setting
$\alpha_c=\alpha_m=\eta_m=\eta_c=0$ and $\chi_c=\chi_m=4\beta_s^2$; in
the thin shell limit, $1-\kappa = d/R_2\ll 1$, we then recover the
well known result $\xi_0= 2 \beta_s \sqrt{3 - 4\beta_s^2}$
\cite{love-elast}. For a nanoshell in a dielectric medium, the
eigenvalues are complex reflecting energy exchanges with the
environment, $\xi=\omega R_2/c_L+i\gamma R_2/c_L$, where
 $\omega=1/T_{osc}$ and $\gamma=1/\tau$ are the mode frequency and 
 damping rate, respectively.

This general model provides an equilibrium description of the
nanoshell acoustic response. However, under ultrafast excitation, the
role of the dielectric core is expected to diminish. Indeed, the
dielectric core is not directly affected by the pump pulse, but
experiences thermal expansion as a result of heat transfer from the
metal shell. At the same time, this expansion is much weaker than that
of the metal, so that when new equilibrium size is established, the
core is almost fully disengaged from the shell. This should be
contrasted to bimetallic particles where the core remains engaged
after the expansion and thus contributes to the acoustical vibration
spectrum \cite{hartland-bimetal1,hartland-bimetal2}. 

\begin{figure}[t]
\begin{center}
 \includegraphics[width=3.0in]{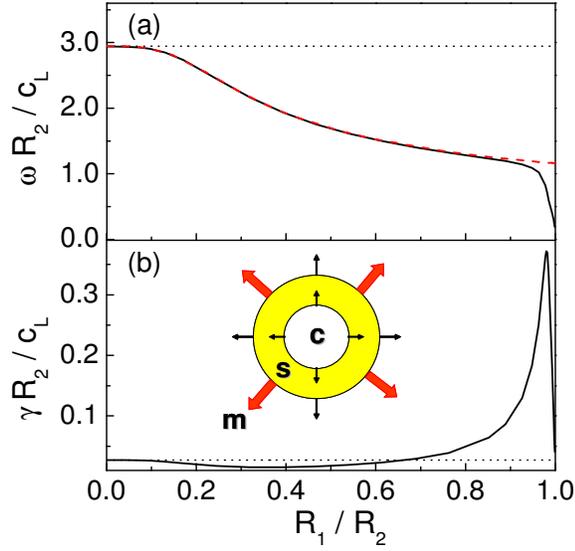}
\end{center}
\caption{\label{fig:spectrum-sym} Calculated frequency (a) and damping
 rate (b)  of the fundamental breathing mode of a gold nanoshell with
disengaged core in water versus its aspect ratio $R_1/R_2$.  The
dashed line in (a) is for a gold nanoshell in vacuum. The horizontal
dotted lines show the normalized frequency and damping for a gold
nanosphere of the same overall size (radius $R_2$). The inset
indicates mechanical movement associated to the fundamental mode (thin
arrows) and the energy damping mechanism to the environment (thick
arrows). } 
 \end{figure}

To take into account this effect, calculations were performed for gold
nanoshells with disengaged core [$\eta_c=0$ in 
Eq.\ (\ref{breathmodes})]. The calculated frequency, $\omega$, and
damping rate, $\gamma=1/\tau$, are plotted in Fig.\ \ref{fig:spectrum-sym} 
versus the aspect ratio $R_1/R_2$, for the fundamental breathing mode
of nanoshells immersed in water.  The data are normalized in units of 
$c_L/R_2$ so that the corresponding curves for solid nanoparticles are
horizontal lines starting at $R_1/R_2=0$. The sound velocities and the
density were taken as $c_L^{(s)} = 3240$ m/s, $c_T^{(s)} = 1200$ m/s,
$\rho^{(s)} = 19700$ kg/m$^3$ for Au , and $c_L^{(m)}= 1490$ m/s, 
$c_T^{(m)} = 0$, and $\rho^{(m)} = 1000$ kg/m$^3$ for water.

The computed frequency of the fundamental mode is significantly
smaller for nanoshells as compared to gold particles of the same
overall size (Fig.\ \ref{fig:spectrum-sym}). It is about 2 times
smaller for $R_1/R_2=0.5$ and further decreases to about a factor of 3
for thin nanoshells.  In contrast, the aspect ratio dependence of the
computed damping rate is non-monotonic. A minimum is reached at
$R_1/R_2=0.4$, followed by a large increase for thin shells that can
be understood on the basis of energy consideration: the deposited
energy is proportional to the nanoshell volume, $V$, while the
efficiency of energy exchange is determined by the surface area,
$A$. Then, the characteristic time of energy transfer from the shell
to the outer medium is $\tau \sim V/Ac^{(m)}_L\propto d/c^{(m)}_L$, as
opposed to the $R/c^{(m)}_L$ dependence for solid particles
\cite{dubrovskiy81}. With further decrease of the nanoshell thickness,
a sharp change in behavior is seen for both $\omega$ and $\gamma$,
indicating a crossover to an overdamped regime 
(Fig.\ \ref{fig:spectrum-sym}). In this thin shell limit, the spectrum
of vibrational modes is mostly determined by the energy exchange with
environment, as shown by the large deviation of the computed frequency
for a water or vacuum environment for $R_1/R_2 \geq 0.9$ 
(Fig.\ \ref{fig:spectrum-sym}). 

The behavior for the thin shell regime can be better analyzed using
approximated analytical solutions for the vibrations frequency and
damping. For $1-\kappa = d/R_2\ll 1$, Eqs. (\ref{breathmodes}) (with
$\eta_c=0$) reduces to
\begin{equation}
\frac{\chi_c}{1-\kappa}
\biggl(
\chi_m -\chi_c + \frac{\alpha_m \eta_m \xi^2}{\alpha_m - i \xi}
\biggr)
=
\biggl(
\chi_m + \frac{\alpha_m \eta_m \xi^2}{\alpha_m - i \xi}
\biggr) \xi_0^2
- \chi_c \xi^2.
\label{simple}
\end{equation}
In the typical case when the metal shell density is much higher than
that of the surrounding medium, i.e.,
$\eta_m=\rho^{(m)}/\rho^{(s)}\ll 1$, Eq.\ (\ref{simple}) further simplifies to
\begin{equation}
x^2 - 1  =
\frac{\alpha_m \eta_m}{\xi_0 (1 - \kappa)}
\Biggl[
\frac{4 \alpha_m \beta_m^2}{\xi_0} - \frac{x^2}{\alpha_m/\xi_0 - i x}
\Biggr],
\label{simple2}
\end{equation}
where $x = \xi/\xi_0$ and we used  $\chi_m-\chi_c = -4 \eta_m
\alpha_m^2\beta_m^2$ and $\chi_m/\chi_c = 1 - \eta_m \beta_m^2$. Two
regimes can now be clearly identified, governed by the ratio
$\eta_m/(1-\kappa)=R_2\rho^{(m)}/d\rho^{(s)} \approx M_m/M_s$, where
$M_s$ is the metal shell mass, and $M_m$ is the mass of outer medium
displaced by the core-shell particle. Explicit expressions can be
obtained for the cases of ``heavy'' and ``light'' shells. For a
``heavy shell'', $M_s\gg M_m$, the complex eigenvalue is given by
%
\begin{eqnarray}
\xi \simeq
\xi_0 - \frac{\lambda}{2}
\biggl[ \frac{\alpha_m+i\xi_0}{(\alpha_m/\xi_0)^2 + 1}- 4 \alpha_m\beta_m^2
\biggr],
\label{xi-heavy}
\end{eqnarray}
where $\xi_0$ is the eigenvalue for a nanoshell in vacuum and
$\lambda= \alpha_m \eta_m/\xi_0 (1 - \kappa)$. In a good 
approximation, the real part is simply $\xi'\simeq \xi_0$, and is thus
independent of the medium or aspect ratio, in agreement with the full
calculation for $R_1 / R_2 \leq 0.9$ 
(Fig.\ \ref{fig:spectrum-sym}). In contrast the imaginary  
part, although small ($\xi'' \ll \xi'$), is only non-zero in presence
of a matrix and thus depends on both. Putting all together, we obtain
in the ``heavy shell'' regime 
\begin{equation}
\label{heavy}
\omega
\simeq
\frac{ 2 c_L^{(s)}\beta_s}{R_2} \sqrt{3 - 4\beta_s^2},
~
\gamma
\simeq
\frac{c_L^{(m)}}{d}
\frac{2\eta_m \beta_s^2 (3-4\beta_s^2)}
{\alpha_m^2 + 4\beta_s^2(3-4\beta_s^2)}.
\end{equation}
As discussed above, here the damping rate is determined by the shell
thickness rather than by the overall size. In the opposite case of a
``light shell'', $M_s\ll M_m$, the eigenvalue is given by
$\xi\simeq 2\alpha_m\beta_m\bigl(\sqrt{1-\beta_m^2} -i \beta_m\bigr)$, yielding
\begin{equation}
\label{light}
\omega \simeq 2c_T^{(m)}/R_2, ~~ \gamma \simeq \omega c_T^{(m)}/c_L^{(m)}. 
\end{equation}
Note that for a light nanoshell in water ($c_T^{(w)}=0$) the limiting
 frequency vanishes. In the crossover region, the nanoshell frequency
 is significantly lower than in vacuum (Fig.\ \ref{fig:spectrum-sym}).

The above theoretical analysis of the nanoshell vibrational modes is
consistent with experimental data. In the aspect ratio of interest,
$R_1/R_2 \approx 0.75$, the fundamental mode period is considerably
longer than for a pure metal particle of same overall size. For the
investigated particles, the aspect ratio $R_1/R_2$ lies in the range
where the normalized frequency $\omega R_2/c_L$ varies weakly so the
period is almost proportional to $R_2$, in agreement with the
experimental data [Fig.~\ref{fig:oscill}(b)]. A deviation from a
simple $R_2$ dependence towards longer $T_{osc}$ is apparent for the
nanoshell with largest aspect ratio (smallest $R_2$), in agreement
with the calculated vibrational modes spectra. However, the measured
period for a nanoshell in Fig.~\ref{fig:oscill}(a), 
$T_{osc}\approx 38$ ps is larger by about a factor of 2 than that
calculated for the {\em ideal} nanoshell. This discrepancy could be
attributed to structural inhomogeneity of the metal shell. Its porous
(``bumpy'') structure with interstices increases the surface to volume
ratio and, thus, moves the vibrational modes towards that of
effectively {\em thinner} nanoshells. Importantly, such structural
defects drive nanoshell acoustical response {\em away} from solid
nanoparticle, as long as the shell is continuous. Note that
clusterization or aggregation processes that effectively break the
shell geometry will results in an increase, as compared to ideal
shell, of the vibration frequency contrary to the experimentally
observed reduction. This specific acoustic response can thus be used
to unambiguously distinguish different nanoobjects produced during
nanoparticle synthesis, such as nanoparticle clusters and nanoshells. 

The computed damping rate $\tau$ is smaller than the experimental one
by almost a factor of 1.5. A similar discrepancy has been reported for
nanosphere colloidal solutions \cite{delfatti-jcp99}. In theoretical
models, computation is made for one nanoparticle with a given mean 
geometry. Damping is then associated to energy transfer to the
surrounding medium and is thus weak in the case of a water matrix. As
a large number of nanoparticles is simultaneously investigated
inhomogeneous damping due to dephasing of the coherently excited
acoustic oscillations of the nanoparticles is thus expected to play a
dominant role \cite{hartland-jcp99,voisin-physica02}. This statistical
effect that reflects the particle size, shape and structure
distribution is also probably at the origin of the sample to sample
fluctuations of the measured damping rate.

In summary, using a time-resolved pump-probe technique, we have
investigated the acoustic vibration of gold nanoshells in colloidal
solution. The results clearly show oscillations with a period in the
40 ps range, much longer than expected for pure gold nanospheres. 
In agreement with our theoretical model, they have been ascribed to
fundamental breathing vibration of the gold nanoshells, whose acoustic
signature is thus observed here for the first time. Note that such low
frequency vibrational modes (in the 1 cm$^{-1}$ range) are extremely
difficult to observe using spontaneous Raman spectroscopy. These
results stress the importance of time-resolved studies of acoustic
vibrational modes as a new and powerful tool for unambiguous
determination of the structure of synthesized nanoobjects via their
specific acoustic properties.

C.G., P.L., N.D.F and F.V. acknowledge financial support by Conseil R\'{e}gional d'Aquitaine.
A.S.K and T.V.S. acknowledge financial support by National Science
Foundation and by National Institute of Health.


\end{document}